\def\mathbi#1{\textbf{\em #1}}
\documentclass[twocolumn,showpacs,preprintnumbers,amsmath,amssymb]{revtex4}

\usepackage{graphicx}
\usepackage{dcolumn}
\usepackage{bm}
\usepackage{multirow}
\usepackage{amsmath}
\usepackage{float}

\usepackage[dvips]{color}
\definecolor{darkgreen}{rgb}{0,.6,0}

\newcommand{\mub}{$\mu_\text{B}$}

\begin{document}
\title{First-principles study of ferroelectricity induced by p-d hybridization in ferrimagnetic NiFe$_2$O$_4$}

\author{Un-Gi Jong$^1$, Chol-Jun Yu$^{1,*}$, Yong-Su Park$^2$, Chong-Suk Ri$^2$}
\affiliation{$^1$Department of Computational Materials Design (CMD), and \\
$^2$Department of Magnetic Materials, Faculty of Materials Science, Kim Il Sung University, \\
Ryongnam-Dong, Taesong-District, Pyongyang, DPR Korea}
\date{\today}

\begin{abstract}
We investigate the ferrimagnetism and ferroelectricity of bulk NiFe$_2$O$_4$ with tetragonal $P4_122$ ~symmetry by means of density functional calculations using generalized gradient approximation + Hubbard $U$ approach. Special attention is paid to finding the most energetically favorable configuration on magnetic ordering and further calculating the reliable spontaneous electric polarization. With the fully optimized crystalline structure of the most stable configuration, the spontaneous polarization is obtained to be 23 $\mu$C/cm$^2$ along the z direction, which originates from the hybridization between the 3d states of the Fe$^{3+}$ cation and the 2p states of oxygen induced by Jahn-Teller effect.
\end{abstract}
\pacs{41.20.Gz, 77.80.-e, 75.85.+t, 31.15.E-}
\maketitle

Magnetic ferroelectrics that possess ferroelectricity together with some form of magnetic order like ferromagnetic, antiferromagnetic, or ferrimagnetic property in the same crystalline phase have attracted much interest due to their fundamental physics and a great promise for applications in future information technology~\cite{schmidt, Spaldin, cheong, scott, ramesh, Eerenstein, Nan}. In particular, improper multiferroics in which a ferroelectric polarization is induced by electronic correlation effects such as non-centrosymmetric spin, charge, and orbital ordering, is currently under intensive study~\cite{picozzi12, benedek, Giovannetti09, fukushima}. This is related to the suggestion that a magneto-electric (ME) coupling in improper multiferroics is expected to be stronger than in proper case, because both dipolar and magnetic orderings share the same physical origin and occur at the same temperature~\cite{picozzi12}.

In the early days of searching improper multiferroics, spin ordering was accepted to be a main driving force for the manifestation of ferroelectric polarization; examples include rare earth manganites $R$MnO$_3$~\cite{Picozzi07,Malashevich,Yamauchi08,PicozziJP09} and $R$Mn$_2$O$_5$~\cite{Giovannetti08,Wang07,Cao} ($R$ = Tb, Ho, Dy, Lu, $\ldots$), and Li-Cu-based oxides LiCu$_2$O$_2$ and LiCuVO$_4$~\cite{Xiang}. These compounds, however, have antiferromagnetic order, which makes their application strongly limited because of zero magnetization (and therefore uncontrollability), and what is worse, they have quite small electric polarization ($\leq$0.1 $\mu$C/cm$^2$). On the other hand, it was found from first-principles and experiments that, in some ferrites including magnetite Fe$_3$O$_4$~\cite{alexe,yamauchi,fukushima1} and rare earth ferrites $R$Fe$_2$O$_4$ ($R$ = Y, Yb or Lu)~\cite{Ikeda,Xiang07,angst08,angst13}, charge ordering or spin-charge ordering induces a spontaneous polarization of remarkably large value: up to 5 $\mu$C/cm$^2$ in monoclinic Fe$_3$O$_4$~\cite{fukushima1} and even tens of $\mu$C/cm$^2$ in hexagonal LuFe$_2$O$_4$~\cite{angst13}, comparable to those of conventional perovskite ABO$_3$-type ferroelectrics as well as proper multiferroics. In these ferrites, moreover, the charge and spin degrees of freedom of electrons can be controlled simultaneously, because they have ferrimagnetic order and thus may have non-zero magnetization. Recently, two primary non-polar distortions like tilting and rotation of oxygen octahedra driven by electronic factors ({\it i.e.,} Jahn-Teller effects) were found to induce the breaking of inversion symmetry and produce a large ferroelectric polarization in the new class of double perovskites AA$'$BB$'$O$_6$; SrTiO$_3$/PbTiO$_3$ superlattice~\cite{Bousquet}, NaLaMnWO$_6$~\cite{picozzi12,fukushima11} with a very large polarization of about 16 $\mu$C/cm$^2$, and Ca$_3$Mn$_2$O$_7$~\cite{benedek} are typical compounds belonged to this materials class. These materials have weak ferromagnetism and thus a significant ME coupling is expected. Despite the great progress in the field of improper multiferroics, the effort to find new mechanism and design new materials is being continued.

In this letter, we present that spinel-type nickel ferrite NiFe$_2$O$_4$ (NFO) is a promising candidate for improper multiferroics with finite magnetization and considerably large polarization, based on the first-principles density functional theory (DFT) calculations. In fact, spinels exhibit several unusual features including magnetoelectricity and multiferroicity~\cite{picozzi12}. From the experiment, it was established that the bulk NFO shows soft ferrimagnetic ordering below relatively high transition temperature of 850 K with a magnetization of 2 \mub~per formula unit (f.u.)~\cite{McCurrie}. In the recent experiment~\cite{Rathore}, moreover, high dielectric permitivity was observed in NFO nanoparticles to be ranged from 60 to 600 at different frequencies and temperatures. This is comparable with the conventional ferroelectrics and therefore implies the presence of spontaneous polarization in the bulk NFO. However, the ferroelectricity of bulk NFO has neither been calculated nor measured yet, though the intensive studies of electronic~\cite{szotek,meinert,Holinsworth}, magnetic~\cite{Cheng13}, and epitaxial strain and magnetoelastic properties~\cite{Fritsch10PRB,Fritsch11APL,Fritsch12_1} have been conducted.

{\it Which is the most stable structure?}--It is well accepted that NFO crystallizes in the completely inverse spinel structure with a face-centered cubic space group $Fd$\={3}$m$ (No. 227), where half of the Fe$^{3+}$ cations occupy the tetrahedral A-sites (8a), the rest of Fe$^{3+}$ and Ni$^{2+}$ cations are equally distributed over the octahedral B-sites (16d), and the sites of oxygen anions are distinguished into O$_1$ and O$_2$. However, it was not clear how Fe$^{3+}$ and Ni$^{2+}$ cations are distributed over B-sites; when distributing Fe$^{3+}$ and Ni$^{2+}$ over B-sites in the $Fd$\={3}$m$ unit cell with the periodic boundary condition rather than in infinite bulk crystal, it might be possible to generate different space groups whose symmetries are lowered from $Fd$\={3}$m$. It is only recent year that, through the Raman spectra measurements, both NFO single crystals~\cite{Ivanov} and thin films~\cite{Iliev} exhibit short range B-site ordering with tetragonal $P4_122$ (No. 91) (or equivalently $P4_322$) symmetry, though the phase with orthorhombic $Imma$ (No. 74) symmetry was not ruled out.
\begin{figure}[!t]
\includegraphics[clip=true,scale=0.23]{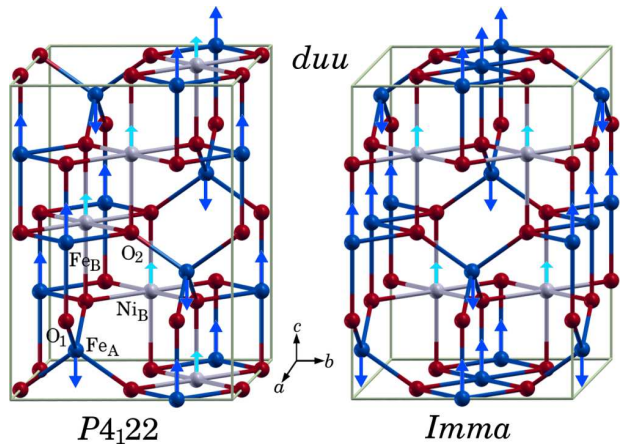}
\caption{\label{fig_conf}(color online) Unit cells with $P4_122$ and $Imma$ symmetries with $duu$ spin configuration of cations Fe$_\text{A}$, Fe$_\text{B}$ and Ni$_\text{B}$, where $u$ and $d$ means spin up and down, respectively.}
\end{figure}

Such experimental finding that NFO has short range ordering was affirmed by the recent first-principles works~\cite{Fritsch10PRB,Fritsch11APL,Fritsch12_1,Cheng13}. Among possible space groups with symmetries lowering from $Fd$\={3}$m$ created by distributing Fe$^{3+}$ and Ni$^{2+}$ cations on 8 B-sites, two relatively high-symmetry space groups $P4_122$ and $Imma$ were found to be energetically the most favorable, and further $P4_122$ slightly lower than $Imma$~\cite{Cheng13}. Therefore, it is only natural that we adopt the tetragonal $P4_122$ and orthorhombic $Imma$ structures as our starting phases of bulk NFO.

Next, we should consider the possible spin configurations of cations, which help to illustrate the nature of the magnetic property of bulk NFO. Denoting spin up and down as $u$ and $d$, there are eight spin configurations of cations, Fe$_\text{A}$, Fe$_\text{B}$ and Ni$_\text{B}$ in due order: $ddd$, $ddu$, $dud$, $duu$, $udd$, $udu$, $uud$, and $uuu$. However, considering that spin inversion will give the same total energy to the origin with only opposite magnetization as confirmed by our calculation, the spin configurations to consider here reduce to four cases: $duu$, $dud$, $uud$, and $uuu$. Fig.~\ref{fig_conf} shows the unit cells of two phases with the $duu$ spin arrangement.

We carried out atomic relaxations for the four spin configurations with $P4_122$ and $Imma$ space groups respectively, using ultrasoft pseudopotential (USPP) plane wave method within spin-polarized DFT as implemented in Quantum ESPRESSO code (version 5.1)~\cite{QE}. The Perdew-Burke-Ernzerhof (PBE) generalized gradient approximation (GGA)~\cite{PBE} was used for the exchange-correlation functional with on-site Coulomb interaction ({\it i.e}. GGA+$U$) to take into account the strong correlation effect of Fe and Ni 3d states~\cite{Liechtenstein}. As in the previous calculations~\cite{Liechtenstein,Cheng13}, the values of $U$ and $J$ for Fe are 4.5 eV and 0.89 eV, and for Ni, different $U$ values ranged from 2 eV to 6 eV, all with $J=1$ eV, were tested. The USPPs are provided in the code, where the valence electrons are taken as 10, 16, and 6 for Ni, Fe and O atoms~\cite{uspp}. The kinetic energy cutoff for plane-wave expansion was set to be 40 Ry and the $k$-points to be ($5\times5\times3$) for the Brillouin zone integration. Marzari-Vanderbilt cold smearing technique~\cite{mvsmearing} with a gaussian spreading factor of 0.02 Ry was applied. We have used tetragonal unit cells containing 4 formula units (28 atoms) with fixed experimental lattice constants $a=b=5.89$ \AA~and $c=8.34$ \AA~\cite{McCurrie}. The convergence threshold of forces on ions was set to be 1.0$\times10^{-4}$ Ry/Bohr.

For all the tested $U_{\text{Ni}}$ values of 2, 4 and 6 eV, the order of four spin configurations in total energy was always like $duu < udu < uud < uuu$ with the almost equal energy difference of 0.2 eV/f.u. between two spin configurations in turn, and this is the same in both $P4_122$ and $Imma$ space groups. Eventually the lowest energy structure was turned out to be $duu$ (or $udd$) configuration for both $P4_122$ and $Imma$. It was also found that $P4_122$ phase is $\sim$0.02 eV/f.u. lower in energy than $Imma$ phase in each spin configuration and this does not change with the  $U_{\text{Ni}}$ value.

Our calculations show that the total magnetization varies with change of $U_{\text{Ni}}$ value; in the case of $duu$ spin configuration, the $U_{\text{Ni}}$ value that gives the total magnetization of 2 $\mu_B$/f.u. consistent with the experimental value for the bulk NFO~\cite{McCurrie}, was 3.24 eV. Therefore, we will use the $U_{\text{Ni}}$ value of 3.24 eV for the following calculations and analysis.

In Table \ref{tab1}, we show the obtained coordinates of atoms and geometries of oxygen octahedra around B-sites cations for the lowest spin configuration $duu$ phases with both $P4_122$ and $Imma$ space groups. While both distortions and tiltings of octahedra due to Jahn-Teller effect are observed in $P4_122$ phase, only slight tiltings are observed in $Imma$ phase.
\begin{table}[!t]
\small
\begin{center}
\caption{\label{tab1} Relaxed atomic coordinates and geometries of oxygen octahedra around B-sites cations for spin configuration $duu$ phases with $P4_122$ and $Imma$ space groups.}
\begin{ruledtabular}
\begin{tabular}{llll|lll}
 & \multicolumn{3}{c|}{$P4_122$} & \multicolumn{3}{c}{$Imma$} \\
 \cline{2-7}
Fe$_\text{A}$ & 4c (x,x,3/8)  & x & 0.2486 & 4e (0,1/4,z)     & z & 0.1282 \\ 
Fe$_\text{B}$ & 4a (0,y,0)    & y & 0.7412 & 4b (0,0,1/2)     &   &        \\ 
Ni$_\text{B}$ & 4b (1/2,y,0)  & y & 0.7435 & 4c (1/4,1/4,1/4) &   &        \\ 
O$_1$         & 8d (x,y,z)    & x & 0.0144 & 8h (0,y,z)       & y & 0.0125 \\ 
              &               & y & 0.7535 &                  & z & 0.7416 \\ 
              &               & z & 0.2423 &                  &   &        \\ 
O$_2$         & 8d (x,y,z)    & x & 0.4904 & 8i (x,1/4,z)     & x & 0.7640 \\ 
              &               & y & 0.7526 &                  & z & 0.5028 \\ 
              &               & z & 0.2452 &                  &   &        \\
\hline
\multicolumn{2}{l}{$\angle$ O$_1$-Fe$_\text{B}$-O$_2$ ($^\circ$)} & \multicolumn{2}{l|}{94.32, 92.05} &   \multicolumn{3}{c}{92.20, 92.17}   \\
 &  &  \multicolumn{2}{l|}{88.50, 85.04} & \multicolumn{3}{c}{87.80, 87.83}        \\
\multicolumn{2}{l}{$\angle$ O$_1$-Ni$_\text{B}$-O$_2$} & \multicolumn{2}{l|}{93.33, 91.89} &   \multicolumn{3}{c}{93.64, 93.61}  \\
 &  &  \multicolumn{2}{l|}{88.72, 86.02} &  \multicolumn{3}{c}{86.36, 86.39} \\
\multicolumn{2}{l}{O$_1$-Fe$_\text{B}$, O$_1$-Ni$_\text{B}$ (\AA)} & \multicolumn{2}{l|}{2.024, 2.047} & \multicolumn{3}{c}{2.017, 2.064} \\
\multicolumn{2}{l}{O$_2$-Fe$_\text{B}$} & \multicolumn{2}{l|}{2.046, 1.997} & \multicolumn{3}{c}{2.026} \\
\multicolumn{2}{l}{O$_2$-Ni$_\text{B}$} & \multicolumn{2}{l|}{2.067, 2.029} & \multicolumn{3}{c}{2.033} \\
\end{tabular}
\end{ruledtabular}
\end{center}
\end{table}
\normalsize

{\it Multiferroicity of bulk NFO}--Here we focus our major interest on the ferroelectricity of NFO because its magnetic property with ferrimagnetic order is well accepted at present~\cite{Cheng13} as also confirmed in this study. Let us consider the aforementioned crystalline symmetries of two phases of bulk NFO. As shown clearly in Fig.~\ref{fig_conf}, the $P4_122$ phase does not have inversion symmetry but the $Imma$ phase has a space inversion center. Therefore, the $P4_122$ phase is allowed to have a spontaneous polarization, while for the $Imma$ phase the spontaneous polarization could not be expected. In addition, the $Imma$ phase can be a centrosymmetric reference phase for the calculation of spontaneous polarization of the $P4_122$ phase.

What leads to whether or not the presence of inversion symmetry is just B-sites ordering of cations; in the former phase it is characterized by $\cdots -B'-B''-\cdots$ chains along [100] and [010] directions, while in the latter case the B-sites ordering is characterized by $\cdots -B'-B'-\cdots$ chains along [100] direction and $\cdots -B''-B''-\cdots$ chains along [010] direction. As pointed out in Ref.~\cite{Ivanov}, the transition from the disordered $Fd$\={3}$m$ phase to the ordered $P4_122$ or $Imma$ phase is of first order and furthermore the phase transition between $P4_122$ and $Imma$ can be occurred by the exchange of Fe$_\text{B}$ in [100] direction and Ni$_\text{B}$ in [010] direction or vice versa.

According to the modern theory of polarization~\cite{Smith,Vanderbilt}, the spontaneous polarization of $P4_122$ NFO is obtained by calculating the polarization difference between ferroelectric $P4_122$ phase and centrosymmetric $Imma$ reference phase, $\Delta \mathbi{P}=\mathbi{P}_{FE}-\mathbi{P}_{Ref}=\Delta \mathbi{P}_{ion}+\Delta \mathbi{P}_{el}$. The ionic contribution $\mathbi{P}_{ion}$ is calculated by the simple ``point charge model'', while the electronic contribution $\mathbi{P}_{el}$ by the Berry phase method as already applied to perovskite ferroelectrics by one of the authors~\cite{yucj07}. The calculations were carried out on the most stable $duu$ spin configurations. The polarizations of $Imma$ NFO for all kinds of spin configurations were calculated to be zero, {\it i.e.,} both ionic and electronic contributions to be zero. In $P4_122$ NFO, the ionic contributions to the polarization were also zero, but the electronic contributions were calculated to be $-23.0$ $\mu$C/cm$^2$ along the $\mathbi{c}$ direction with the polarization quantum modulo $e|\mathbi{c}|/\Omega=46.0$ $\mu$C/cm$^2$ ($|\mathbi{c}|=8.34$ \AA, and $\Omega=289.33$ \AA$^3$). This indicates the $P4_122$ NFO to be improper multiferroics (with the ferrimagnetic order as discussed below). When comparing with the conventional improper multiferroics, the calculated polarization is remarkably large, but not very surprising as compared to other recently found spinel ferrites like Fe$_3$O$_4$, LuFe$_2$O$_4$ and NaLaMnWO$_6$~\cite{Xiang07,angst13,fukushima1,fukushima11}. We note that the polarization comes from spin down electrons for the case of $duu$ spin configuration, while for the case of $udd$ spin configuration (another lowest-energy spin configuration with the opposite magnetic moment to the $duu$ configuration) it is from spin up electrons. This feature is very precious for spintronic applications because it gives the way to control the only electrons with a certain spin direction by applying external electric field.

\begin{figure}[!t]
\begin{center}
\includegraphics[clip=true,scale=0.35]{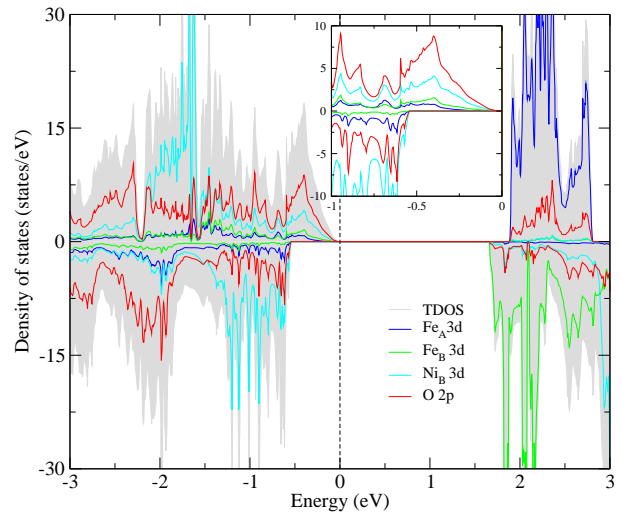}
\end{center}
\caption{\label{fig_dos}(color online) Density of states of NiFe$_2$O$_4$ with $P4_122$ space group and $duu$ spin configuration. $E_\text{F}$ is set to be zero.}
\end{figure}

To see an electronic essence behind having large polarization, we resort to the electronic properties of bulk NFO. The plots of density of states (DOS) (Fig.~\ref{fig_dos}) calculated for $P4_122$ NFO with $duu$ spin configuration show the band gaps of 1.88 eV for spin up electrons and 2.30 eV for spin down ones, which are in good agreement with the recent theoretical and experimental values~\cite{meinert}. The atom-projected partial DOS indicates that the 3d states of cations hybridize strongly with the O 2p states for both spin up and down electrons. Moreover, we can see that below Fermi level the Ni$_\text{B}$ 3d electrons play the major role for both spin up and down, while above Fermi level the Fe$_\text{A}$ 3d electrons for spin up and the Fe$_\text{B}$ 3d electrons for spin down play the dominant role.

To illustrate how $P4_122$ NFO has large polarization, we calculated maximally localized Wannier functions (MLWF) of valence bands for spin up and down states in the $duu$ spin configuration, and recalculated the polarization by summing up the displacements of MLWF centers $\Delta\mathbi{r}$ using the following equation, \[\mathbi{P}_{el}=-\frac{e}{\Omega}\sum_{m}({\Delta\mathbi{r}}_{m,\uparrow}+{\Delta\mathbi{r}}_{m,\downarrow}),\]
where $m$ runs over MLWF centers (with a charge $-e$ located at $\mathbi{r}_m$) in the unit cell with the volume $\Omega$ and arrows denote the spin states. In accordance with the Berry phase calculation, the sum of the displacements of MLWF centers for spin up state was calculated to be zero in all Cartesian directions ({\it i.e.,} $\sum_{m}\Delta\alpha_{m,\uparrow}=0; \alpha=x, y, z$), while one for spin down states was to be $4.16$ \AA~in the $z$ direction with also zero values in the $x, y$ directions ({\it i.e.,} $\sum_{m}\Delta x_{m,\downarrow}=\sum_{m}\Delta y_{m,\downarrow}=0, \sum_{m}\Delta z_{m,\downarrow}=4.16$ \AA). The polarization calculated by the above formula with the relevant variables (the sum of the displacements $\sum_{m}\Delta z_{m,\downarrow}=4.16$ \AA~and the unit cell volume $\Omega=289.33$ \AA$^3$) gives exactly the same value as the Berry phase calculation.

{\it What causes the ferroelectric polarization?}--To clarify the mechanism that leads to such a large ferroelectric polarization in bulk NFO, we made a careful analysis of the electronic properties including removal of degeneracy in B-sites cations through Jahn-Teller effect and hybridization between 3d states of Fe$^{3+}$ cations on B-sites and 2p states of oxygens.

\begin{figure}[!ht]
\includegraphics[clip=true,scale=0.32]{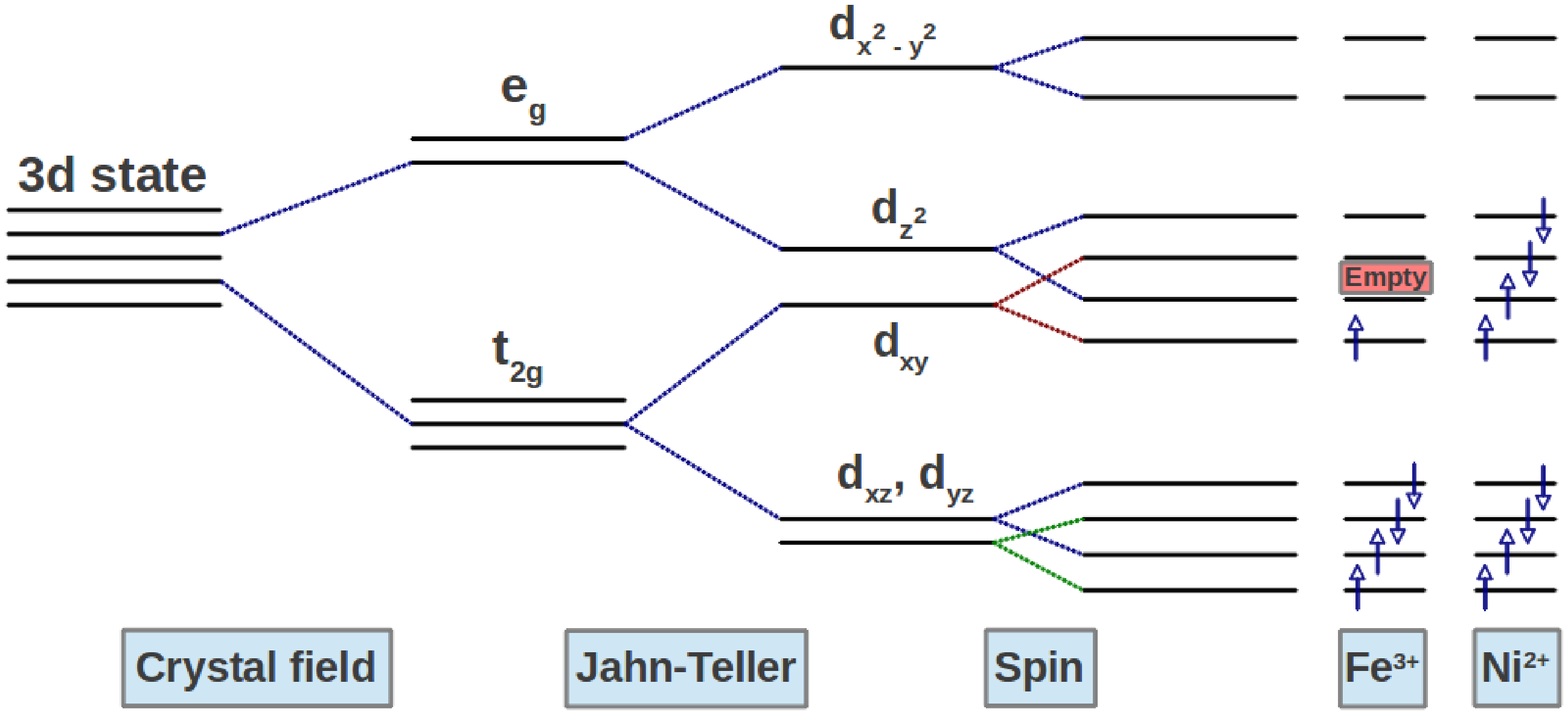}
\caption{\label{fig_level}(color online) Schematic view of Fe$^{3+}$ and Ni$^{2+}$ energy levels showing the removal of degeneracy}
\end{figure}

As mentioned above, the key difference between ferroelectric $P4_122$ phase and centrosymmetric $Imma$ phase is the B-sites ordering of cations. While in both phases the 3d states of cations were lifted into $\it{e_g}$ and $\it{t_{2g}}$ states owing to the crystal field effects, the $\cdots -B'-B''-\cdots$ chains along [100] and [010] directions in the case of $P4_122$ phase, being similar to the double perovskites, induce another effect, $\it{i.e.}$, Jahn-Teller effect, which lifts the double degeneracy of $\it{e_g}$ state and the triple degeneracy of $\it{t_{2g}}$ state. In the context of atomistic structure, Jahn-Teller effects cause slight distortions and tiltings of the oxygen octahedra in the relaxed $P4_122$ phase, breaking their local symmetries. However, the ionic contributions to the polarization by locally distorted and tilted octahedra add up to zero, reflecting the fact that total ionic symmetry remains centrosymmetric during the structural relaxation. In addition to the crystal field and Jahn-Teller effects, spin degree of freedom gives rise to the spin splitting. In Fig.~\ref{fig_level}, we show the lifting of energy levels of Fe$^{3+}$ and Ni$^{2+}$ cations by such effects. These energy level splittings can be shown from the band structure and partial DOS for spin up and down channels.

\begin{figure}[!t]
\includegraphics[clip=true,scale=0.23]{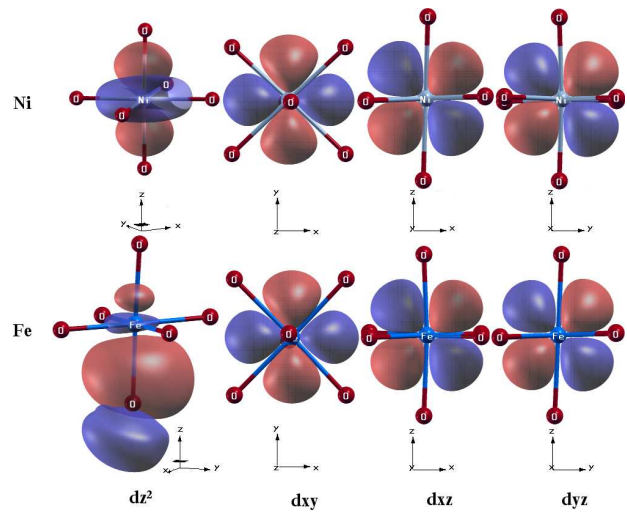}
\caption{\label{fig_wan}(color online) Isosurface plots of maximally localized Wannier functions associated with the 3d states of Fe$^{3+}$ and Ni$^{2+}$ cations for spin down case. The isosurface value is $\pm 1/\sqrt{\Omega}$, where $\Omega$ is the unit-cell volume.}
\end{figure}

The calculated MLWFs pointed towards the importance of the lowest empty d$_{z^2}$ state of spin down channel in conduction bands of Fe$^{3+}$ cation, introducing the hybridization with the p$_z$ state of oxygen. This p-d hybridization leads to breaking of the inversion symmetry driven by the electronic degree of freedom. In order to illustrate how the p-d hybridization breaks the inversion symmetry, we display the MLWFs associated with the 3d states of Fe$^{3+}$ and Ni$^{2+}$ cations for spin down case in the $duu$ configuration. As shown in Fig.~\ref{fig_wan}, the hybridization between the d$_{z^2}$ state of the Fe$^{3+}$ cation on B-site and the 2p state of the oxygen is clearly proved, showing the character of the p$_z$ of oxygen and the d$_{z^2}$ of the Fe$^{3+}$ cation. Here, we place great emphasis on the fact that the ferroelectric property is originated from the p-d hybridization induced by Jahn-Teller effect, and the magnetic property also comes from the 3d states of cations. We note that such properties give rise to the valuable applications for magnetoelectric actuators, spintronics and so on, since they are originated from the same electronic degree of freedom.

In conclusion, we have investigated the ferroelectricity induced by p-d hybridization in ferrimagnetic NiFe$_2$O$_4$ with P4$_122$ space group symmetry by means of GGA$+U$ method. The polarization of 23 $\mu$C/cm$^2$ is considerably large, compared to other improper multiferroics, and the net magnetic moment of 2 $\mu_\mathrm{B}$/f.u. is also large. This work sheds light on that the driving force of the large ferroelectric polarization is just the hybridization between the empty $3d$-states of Fe$^{3+}$ cations on B-sites and the $2p$-states of the oxygen anions, induced by Jahn-Teller effect. Although in ferrimagnetic NiFe$_2$O$_4$ with P4$_122$ space group such a large ferroelectric polarization was not yet measured experimentally, we believe that our findings will be confirmed by further experimental and theoretical studies.

We thank Stefaan Cottenier at the Center for Molecular Modeling (CMM) $\&$ Department of Materials Science and Engineering (DMSE), Ghent University, for his careful reading of the manuscript and invaluable comments. This work was supported partially by the Committee of Education, DPR Korea under the project entitled ``Strong correlation phenomena at superhard, superconducting and nano materials''.

\end{document}